\documentclass[prl, aps, superscriptaddress, twocolumn, showpacs]{revtex4}
\usepackage[cp1250]{inputenc}
\usepackage[T1]{fontenc}

\usepackage{graphicx}

\begin{document}

\newcommand{\rmi}[1]{_{\mathrm{#1}}}

\title{Nonlinear Faraday Rotation and Superposition-State Detection in Cold Atoms}

\author{Adam Wojciechowski}
\affiliation{Institute of Physics, Jagiellonian University,
Reymonta 4, PL-30-059 Krak\'ow, Poland} \affiliation {Joint
Krakow-Berkeley Atomic Physics and Photonics Laboratory, Reymonta
4, PL-30-059 Krak\'ow, Poland}

\author{Eric Corsini}
\affiliation{Department of Physics, University of California,
Berkeley, CA 94720-7300, USA} \affiliation {Joint Krakow-Berkeley
Atomic Physics and Photonics Laboratory, Reymonta 4, PL-30-059
Krak\'ow, Poland}
\author{Jerzy Zachorowski}
\affiliation{Institute of Physics, Jagiellonian University,
Reymonta 4, PL-30-059 Krak\'ow, Poland}\affiliation {Joint
Krakow-Berkeley Atomic Physics and Photonics Laboratory, Reymonta
4, PL-30-059 Krak\'ow, Poland}

\author{Wojciech Gawlik}
\affiliation{Institute of Physics, Jagiellonian University,
Reymonta 4, PL-30-059 Krak\'ow, Poland}\affiliation {Joint
Krakow-Berkeley Atomic Physics and Photonics Laboratory, Reymonta
4, PL-30-059 Krak\'ow, Poland}

\date{\today}

\begin{abstract}
We report on the first observation of nonlinear Faraday rotation
with cold atoms at a temperature of $\sim$100 $\mu$K. The observed
nonlinear rotation of the light polarization plane is up to 0.1
rad over the 1~mm size atomic cloud in approximately 10~mG
magnetic field. The nonlinearity of rotation results from
long-lived coherence of ground-state Zeeman sublevels created by a
near-resonant light. The method allows for creation, detection and
control of atomic superposition states. It also allows
applications for precision magnetometry with high spatial and
temporal resolution.
\end{abstract}

\pacs
{33.57.+c, 42.50.Dv, 42.50.Gy, 32.80.Xx}

\keywords{Nonlinear Faraday Effect, quantum superpositions,
Magnetooptical trap, MOT}

\maketitle

The linear Faraday rotation (LFR) of the polarization plane of
light propagating in the medium is a well known consequence of
optical anisotropy caused by a longitudinal magnetic field. For
thermal gases the Doppler effect broadens the range of the
magnetic fields where the effect is visible and reduces the size
of the maximum rotation relative to atoms at rest. The use of cold
atoms with their Doppler width narrower than the natural linewidth
distinguishes this situation from experiments at room temperature.
The experiments on LFR with cold atoms were performed in a
magneto-optical trap (MOT) \cite{FrankeArnold2001, Labeyrie2001,
Narducci2003}, and in an optical dipole trap
\cite{Terraciano2008a}.

Application of strong, near-resonant laser light may result in
the creation of coherent superpositions of Zeeman sublevels of an
atomic ground state. Such superpositions (Zeeman coherences) are
known to be responsible for a variety of coherent phenomena in
light-matter interaction, like coherent population trapping
\cite{Arimondo1996}, electromagnetically-induced transparency
\cite{Fleischhauer2005}, nonlinear magneto-optical rotation or
nonlinear Faraday rotation (NFR) \cite{Budker2002b} and their
interplay \cite{Drampyan2009}. Superposition states are also at
the heart of quantum-state engineering (QSE). Most of QSE
experiments require {initial states of well defined atomic spin
(or total angular momentum $F$), usually prepared in a stretched
state, which is realized by putting most of (ideally all) atomic
population into a Zeeman sublevel with extreme value of magnetic
quantum number $m$ \cite{Julsgaard2004}. Below we report how
superpositions of specific Zeeman sublevels, or Zeeman coherences
belonging to a given $F$ are created in cold ($\sim$100~$\mu$K)
atomic samples and observed with high sensitivity using nonlinear Faraday
rotation. In the experiment laser light both creates and detects
the Zeeman coherences. The same detection technique can be applied
to detect the presence of Zeeman coherences already introduced
with other mechanisms. Furthermore, the time-dependent
detection provides information on the temporal evolution of the
superposition states.

The described experiment shows the potential of NFR with cold
atoms for precision magnetometry with prospective $\mu$G
sensitivity, large dynamic range (zero-field to several G), and
sub-mm spatial resolution in magnetic field mapping. Magnetic
field sensing with cold atoms utilizing Larmor precession
of alkali atoms in a magnetic field has been discussed in: MOT
\cite{Isayama1999}, Bose-Einstein condensate \cite{Wildermuth2006,
Vengalattore2007} and an optical dipole trap
\cite{Terraciano2008b}. Our measurements apply a different
principle: rather than measuring Larmor frequency (single atom
quantity), we measure rotation of a polarization plane (a
cumulative effect over the whole sample), which may offer higher
accuracy in very low magnetic fields. In our experiment the
rotation is mainly caused by the nonlinear medium's birefringence
resulting from the light-induced Zeeman coherences
\cite{Series1966, Cohen1967}, regarded as the diamagnetic effect.
The rotation resulting from population imbalance (paramagnetic
effect) was studied with cold atoms in recent experiments devoted
to spin squeezing \cite{Kubasik2009}.

For resonant excitation, rotation angle $\theta$ is a measure of
circular birefringence, $\theta \propto (n_+ - n_-)$, where
$n_\pm$ are the refractive indices for  $\sigma^{\pm}$ polarized
light and $n_\pm - 1 \propto \mathcal{E}^{-1}
\sum_{eg}\,Re(d_{eg}\rho_{eg})$ with $\mathcal{E}$ being the light
electric field amplitude, $d_{eg}$ the dipole moment, and
$\rho_{eg}$ the density matrix element. The summation goes over
all ground- and excited-state sublevels $g$ and $e$ linked by the
allowed transitions, as shown in Fig. \ref{fig:setup+levels}b. In
the stationary regime, $\rho_{eg}$ can be expressed as
$\rho_{eg}=\sum_{e'g'}
(\Omega_{eg'}\rho_{g'g}-\rho_{ee'}\Omega_{e'g})/(\delta_{eg}-i\Gamma/2)$,
where $\delta_{\alpha\beta}$ and $\Omega_{\alpha\beta}$ denote
respectively the light detuning and Rabi frequency for the
$\alpha\leftrightarrow\beta$ transition, and $\Gamma/2$ is the
relaxation rate of the optical coherence. This relation indicates
that optical coherences, and consequently also the refractive
indices and rotation angle, depend on the density matrix elements
$\rho_{g'g}$ and $\rho_{ee'}$ which
represent populations of and coherences between Zeeman sublevels
of the ground and excited states. For not-too-strong
light, the excited-state coherences are negligible and
all couplings shown in Fig. \ref{fig:setup+levels}b form
independent generic $\Lambda$-systems which involve coherences
between ground-state sublevels with $\Delta m= \pm2$.

The main difficulty in observation of NFR with cold atoms is that
at light intensity required for creation of the Zeeman coherence
the laser beam may mechanically perturb the cold-atom sample. In our study
this adverse effect is reduced  by retroreflection of the light
beam and careful optimization of the experimental conditions to
minimize the light power.

\begin{figure}
  \includegraphics[width=\columnwidth]{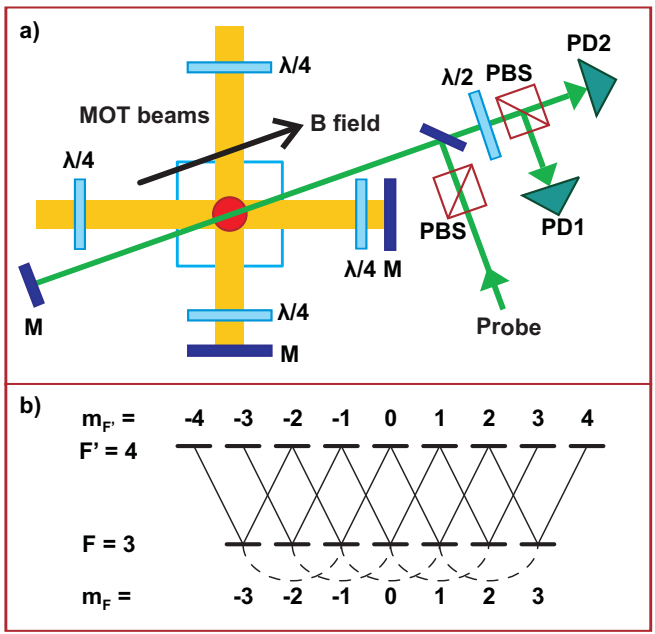}\\
  \caption{(a) The setup of the experiment for the balanced polarimeter arrangement.
  M are mirrors, PBS polarizing beam splitters, PD photodetectors,
  $\lambda/2$, $\lambda/4$ waveplates. Direction of the magnetic field $B$
  necessary for the observation of the Faraday rotation is indicated.
  (For FS scheme the $\lambda/2$ plate is removed and PD2 is not used).
  (b) Energy level-structure with the Zeeman-coherences established by a linearly-polarized light.}
  \label{fig:setup+levels}
\end{figure}

The experiment (see the setup shown in
Fig. \ref{fig:setup+levels}a) was performed with about $10^{7}$ $^{85}$Rb atoms
using a standard MOT. In addition to the trapping and repumping
lasers we used a separate probe laser whose frequency was tuned
around the $F=3 \rightarrow
F'=4$ hyperfine transition of the D2 line (780nm).
Fig. \ref{fig:setup+levels}b depicts the Zeeman structure of the $F=3$ and
$F'=4$ states with the transitions induced by linearly polarized
light (superposition of $\sigma^{\pm}$ polarizations).
A weak linearly
polarized probe beam of several $\mu$W in power and 2~mm in
diameter was sent through the atom cloud and then retroreflected
to partially reduce light pressure effects. The probe-beam frequency
14~MHz below the line center proved to be optimal from the point
of view of atomic loss which we attribute to extra Doppler-cooling
mechanism by two counter-propagating beams. The double passage of
light through the sample doubled the acquired Faraday rotation.
The light polarization was measured in two arrangements: using
balanced polarimeter (direct rotation angle measurement) and in a
crossed polarizers or forward-scattering (FS) scheme which
for resonant light is sensitive to the square of the
rotation angle. For non-resonant case, circular dichroism
contributes also to the observed signal.

In the experiment atoms were collected and cooled in the MOT. This
phase was periodically interrupted for the measurement of optical
rotation: the MOT lasers and the quadrupole magnetic field were
switched off and a homogenous magnetic field $B$ of a controlled
value was applied along the probe beam. After 2~ms (required for
complete decay of the eddy currents induced by turning off the
quadrupole field) the probe beam was switched on and polarization
rotation was recorded for the next 5~ms. Finally, the MOT fields
were switched back for 50--200~ms and the atomic cloud was
recaptured and cooled. This procedure allowed recording polarization
rotation signals as a function of time for each value of the B
field. The experiment was controlled by a PC, which also
digitized, stored, and averaged (typically 20 times) the data.

\begin{figure}[t]
  \includegraphics[width=\columnwidth]{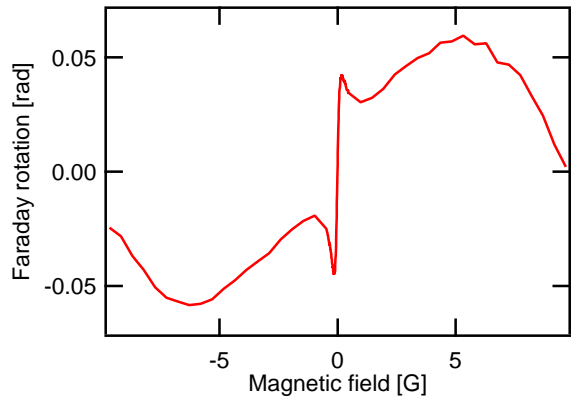}\\
  \caption{Linear (wide) and nonlinear (narrow) Faraday rotation resonances centered at $B=0$.
  NFR resonance is power broadened for a better visibility. The probe power is 64 $\mu$W.}
  \label{fig:Fig2}
\end{figure}

Typical signals (rotation angle vs. $B$) associated with
linear and nonlinear Faraday effect have the form of dispersive
resonances nested at $B=0$, as shown in Fig. \ref{fig:Fig2}. The
narrow feature is the nonlinear resonance (NFR); it appears when
the probe beam is sufficiently intense. Hereinafter, we refer to
this nonlinear resonance as the zero-field resonance. The width of
the linear resonance amounts to several G and corresponds to the
natural linewidth of the studied transition. It also depends on
the detuning of the probe beam from resonance condition and
initial Zeeman-sublevel populations, as has been shown in
\cite{Labeyrie2001}. That situation is prominently different from
the case of vapor cells, where LFR resonance is two orders of
magnitude broader,
because of the Doppler effect

\begin{figure}[t]
  \includegraphics[width=\columnwidth]{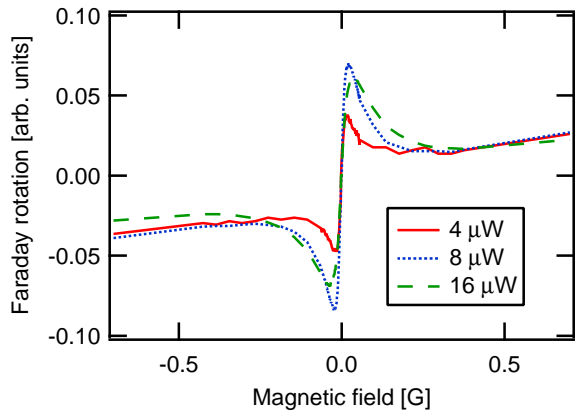}\\
  \caption{Evolution of the Faraday rotation with increasing
  light intensity of the probe beam showing the nonlinear increase
  and power broadening of the central resonance.}\label{fig:Fig3alt}
\end{figure}

In Fig. \ref{fig:Fig3alt} we depict the evolution of the Faraday
rotation signals with the increase of the light power. While the
wide structure associated with the linear Faraday effect represent
rotation angle independent on the light intensity, the central
narrow feature clearly exhibits nonlinear behavior. The narrow
part is due to the superpositions of the ground-state Zeeman
sublevels which differ by $|\Delta m|=2$, shown in Fig.
\ref{fig:setup+levels}b. These are thus the light-induced Zeeman
coherences that are responsible for nonlinearity of the Faraday
effect observed with appropriately strong light in very small
magnetic fields. The narrow width reflects the long lifetime of
the ground-state superpositions which is a necessary prerequisite
for qubits and QSE applications. In case
of atoms released from the MOT, the main mechanism of the
resonance broadening is the escape time of atoms from the
observation volume due to gravitation and their initial momenta.
There is also light-induced expelling of atoms from the probed
volume which can be seen in Fig. \ref{fig:Fig3alt} as the drop of
maximal rotation seen with 16 $\mu$W relative to 8 $\mu$W. Another
major contribution comes from transverse magnetic fields, and can
be understood as power broadening due to magnetically-driven
transitions between degenerate Zeeman sublevels for the near-zero
fields. Therefore, DC and low-frequency transverse magnetic field
components have to be precisely compensated for the NFR
observation. Other broadening mechanisms include gradient of the
longitudinal magnetic field and power broadening due to the
probing beam, which can be reduced by using appropriate intensity
and detuning. The latter offers additional possibility of laser
cooling for retroreflected, red detuned probe beam, as mentioned
above.

\begin{figure}
  \includegraphics[width=0.9\columnwidth]{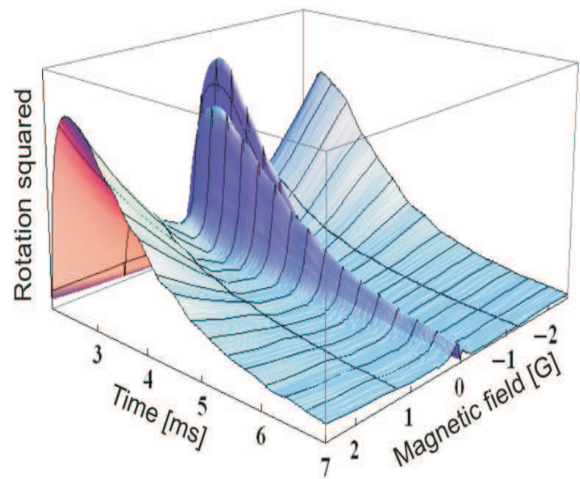}\\
  \caption{Time evolution of linear (wide) and nonlinear (narrow) Faraday rotation resonances in FS-arrangement where the signal is proportional to magneto-optical rotation angle squared. Probe power is 18~$\mu$W.}
  \label{fig:3d}
\end{figure}

Time evolution of Faraday rotation squared (FS-arrangement) is
depicted in Fig. \ref{fig:3d}. The linear effect depends only on
number of atoms and Zeeman population distribution and thus
follows the temporal evolution of these quantities. The nonlinear
effect results from light-atom interaction, i.e., optical pumping
with linearly polarized light and thus requires some,
light-intensity-dependent, time to build up. This effect can be
seen in Fig. \ref{fig:Fig5a}, where examples of such evolution for
two different light intensities are presented. Unlike the linear
contribution, the onset of which is limited only by the detector
time constant, the initial slopes of the nonlinear contributions
indeed depend on the probe power. Both signals decay as atoms
escape from the probed volume. The separation between the
linear and nonlinear contributions corresponding to a given light
power is done based on the magnetic field strength: at about 45~mG
the LFR is negligible and NFR dominates the rotation, whereas the
opposite is true for magnetic fields of 3~G and above.

\begin{figure}[t]
  \includegraphics[width=\columnwidth]{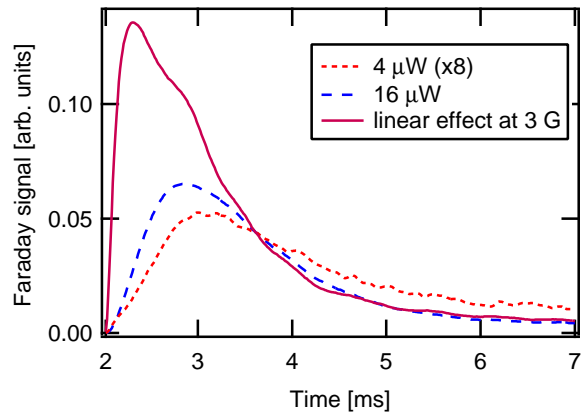}\\
  \caption{Time dependence of the nonlinear Faraday rotation with a 45~mG magnetic field at two different probe beam powers (4~$\mu$W - magnified 8 times and 16~$\mu$W) compared to the time dependence of the linear Faraday rotation at 3~G and 16 $\mu$W.}
  \label{fig:Fig5a}
\end{figure}

Since the stationary ground-state coherences are destroyed when
Larmor precession becomes faster than the coherence relaxation
time, direct observation of the NFR signals is limited to a very
narrow (some mG) range around $B=0$. One possibility to observe
NFR not only around the zero magnetic field is to use modulation
techniques. Two arrangements have been proposed using either
frequency (FM NMOR \cite{Budker2002a}) or amplitude (AMOR
\cite{Gawlik2006}) modulation of light. In both arrangements
strobed pumping creates the modulated Zeeman coherence
and phase sensitive detection is used to extract the
magneto-optical rotation amplitude. In addition to the zero-field
resonance, two other resonances appear in the demodulated rotation
signal when the modulation frequency $\Omega_{m}$ meets $\pm$
twice Larmor precession frequency in a given magnetic field. These
high-field resonances result from the optical pumping synchronous
with the Larmor precession. The factor of 2 appears because the
two-fold symmetry of the optical anisotropy associated with
$|\Delta m| =2$ coherences yields modulation at precisely twice
the Larmor precession. The width of these resonances is determined
by the coherence lifetime and, in case of long-lived ground
states, can be as narrow as the zero field resonance.

In our experiment the AMOR technique was applied: the
probe beam was periodically chopped using the acousto-optical modulators. Use of modulation frequencies up
to $\sim$10~MHz allowed detection of resonances in
magnetic fields as large as 9~G. This is an order of magnitude
higher field compared to previous FM NMOR and AMOR work and
demonstrates the method's potential for precision magnetometry in
a wide range of fields. This range can be further extended by
using electro-optical modulators up to the fields where the
nonlinear Zeeman effect starts to affect the signals. Figure
\ref{fig:Fig6} shows NFR signal with two AMOR resonances at $\pm
3$~G that are the evidence of driving $|\Delta m| = 2$ coherences
at non-zero magnetic fields.

\begin{figure}[t]
  \includegraphics[width=\columnwidth]{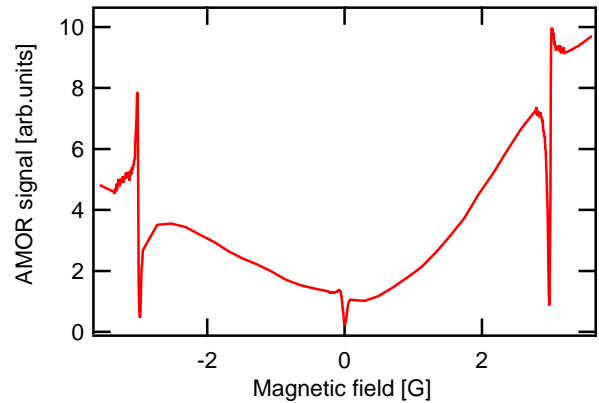}\\
  \caption{NFR with amplitude modulated light (AMOR). The narrow central resonance is a typical NFR zero-field resonance and the two high field resonances at $\pm 3$~G result from amplitude modulation of the light with $\Omega_{m}=2.8$~MHz. Presence of such high-field resonances allows for precision magnetometry of non-zero magnetic fields. The broad background is the LFR. The slight asymmetry of resonance shapes can be attributed to experimental setup imperfection.}
  \label{fig:Fig6}
\end{figure}

In conclusion, we have demonstrated the nonlinear Faraday rotation
for a sample of cold atoms both with cw and modulated laser beams.
The use of retroreflected beam alleviated the problem of
mechanical perturbation of the cold atoms by the probe beam. In
contrast to previous experiments with pure quantum states of
oriented spins, the NFR measurements allow control and convenient
studies of long-lived superposition states of aligned spins, i.e.
quantum superpositions of Zeeman sublevels belonging to a given
$F$. In particular, we are able to vary the degree of Zeeman
coherence and monitor its build-up and decay, both in the
stationary regime ($B\simeq0$), and for the Larmor frequencies up
to 10~MHz. In addition to its potential for QSE, the NFR effect
can be used for measuring a wide range of transient and static
magnetic fields with 10~$\mu$s time resolution, sub-mG
sensitivity, and mm spatial resolution given by the size of the cold atom cloud or the beam waist size. The current results are limited
mostly by finite lifetime of trapped atoms and power broadening by
the probe beam. Transfer of atoms into an optical dipole trap
would make probing time much longer ($\sim$1~s) and the light-atom
coupling more effective whereas the use of separate pump and probe
beams as opposed to a single pump-probe beam would alleviate power
broadening limitations.

The authors would like to acknowledge valuable discusions with D.~Budker, 
W.~Chalupczak, R.~Kaiser, M.~Kubasik and S.~Pustelny. This work has been supported 
by the Polish Ministry of Science (grant \# N N505 092033 and N N202 046337) 
and NSF Global Scientists Program.

\end{document}